\documentclass[aps,pra,twocolumn,showpacs]{revtex4}

\usepackage{bm}
\usepackage{epsfig}
\usepackage{psfig}

\newcommand{\lm}{{\cal P}}

\begin{document}

\title{Solitons, solitonic vortices, and vortex rings in a
cylindrical Bose-Einstein condensate}
\author{S. Komineas$^1$ and N. Papanicolaou$^2$}
\affiliation{$^1$Theory of Condensed Matter Group,  Cavendish Laboratory,
Madingley Road, Cambridge CB3 0HE United Kingdom \\
$^2$Department of Physics, University of Crete, and Research Center of Crete,
Heraklion, Greece}

\date{\today}

\begin{abstract}
Quasi-one-dimensional solitons that may occur in an elongated
Bose-Einstein condensate become unstable at high particle density.
We study two basic modes of instability and the corresponding
bifurcations to genuinely three-dimensional solitary waves
such as axisymmetric vortex rings and non-axisymmetric solitonic
vortices.
We calculate the profiles of the above structures and examine their
dependence on the velocity of propagation along a cylindrical trap.
At sufficiently high velocity, both the vortex ring and the solitonic vortex
transform into an axisymmetric soliton. We also calculate the energy-momentum
dispersions and show that a Lieb-type mode appears in the excitation
spectrum for all particle densities.
\end{abstract}

\maketitle

\section{Introduction}

The possible occurrence of solitons in a Bose-Einstein condensate (BEC)
was theoretically predicted sometime ago, within a one-dimensional (1D)
Gross-Pitaevskii model \cite{tsuzuki,zakharov},
but experimental observation was delayed because of the absence of a physical
realization of a strictly 1D Bose gas.
Nonetheless, similar coherent structures were recently observed in BECs
confined in traps of varying geometry \cite{burger,denschlag}
by a method (phase imprinting) that was directly inspired by the
analytical structure of 1D solitons. Yet it has become clear that
quasi-1D solitons are susceptible to various instabilities  within
the three-dimensional (3D) environment of realistic traps.

Indeed, a stability analysis based on the linear Bogoliubov-de Gennes (BdG)
equations \cite{muryshev1} revealed that an axisymmetric soliton is
stable only at sufficiently low particle number or high aspect ratio
in an elongated trap. The primary mode of instability was shown to result
from non-axisymmetric perturbations, in the sense that a purely imaginary
eigenvalue first appears in the spectrum of the BdG equations with azimuthal
angular momentum $m\!=\!1$. This ''snake instability'' was further
analyzed in Ref.~\cite{feder} and argued to be responsible for a possible
decay of a soliton into vortex rings and/or vortices. In fact, based on the
above theoretical work, non-stationary vortex rings were experimentally
detected in a spherical trap  \cite{anderson}. On the other hand,
Brand and Reinhardt \cite{brand1,brand2} suggested that the primitive
mode associated with the snake instability is a stable structure
that may be called solitonic vortex.

The emerging picture is sufficiently perplexing to deserve closer attention.
In some recent work \cite{komineas1,komineas2} we have carried out a
detailed calculation of axisymmetric solitons in a cylindrical trap,
which were shown to become unstable at high particle density even if
we restrict ourselves to radial ($m\!=\!0$) perturbations. The soliton was then
shown to bifurcate to an axisymmetric vortex ring with lower energy.
Here, we complete this work by allowing azimuthal deformations of the
soliton, in order to eventually account for the primary ($m\!=\!1$)
instability discussed in Ref.~\cite{muryshev1,feder}.
We thus obtain a reasonably complete description of the two basic modes
of instability of the soliton and the corresponding bifurcations to
axisymmetric vortex rings and non-axisymmetric solitonic vortices.

The simplest theoretical picture is obtained in the ideal limit of an
infinitely elongated cylindrical trap, because static solitary waves
may then be treated on equal footing with those propagating at nonzero
velocity. Hence, all calculations presented in the main text pertain
to an infinite cylindrical trap. In Sec.~II we formulate the problem
and briefly describe the numerical methods employed. The three basic
types of static solitary waves (solitons, solitonic vortices, and
vortex rings) and their relative stability are discussed in Sec.~III,
whereas solitary waves moving along the cylindrical trap at constant
velocity are calculated in Sec.~IV. The main conclusions are summarized
in Sec.~V, and some important technical details are relegated to two
appendices.
In particular, the case of a finite axisymmetric trap is discussed in
Appendix B.

\section{formulation}
An axisymmetric harmonic trap is characterized by a transverse confinement
frequency $\omega_\perp$ and a longitudinal frequency $\omega_\|$. We first
consider the special limit of a cylindrical trap, with $\omega_\|\!=\!0$,
a restriction that is relaxed only in Appendix B. In this limit,
physical units are rationalized as in Ref.~\cite{komineas2}.
Thus time is measured in units of $1/\omega_\perp$ and distance in units
of the transverse oscillator length $a_\perp\!=\!(\hbar/M \omega_\perp)^{1/2}$
where $M$ is the mass of each atom. Complete specification of the system
also requires as input the linear density $\nu$ which is equal to the
average number of particles per unit length of the cylindrical trap.
An important dimensionless parameter is then given by
$\gamma\!=\! \nu a$ where $a$ is the scattering length. A field rescaled
according to $\Psi \to \sqrt{\nu} \Psi/a_\perp$ satisfies the rationalized
Gross-Pitaevskii equation
\begin{equation}
\label{eq:1}
 i \frac{\partial\Psi}{\partial t}  =  -\frac{1}{2} \Delta\Psi
 + V_{\rm tr} \Psi + g |\Psi|^2 \Psi\,,
\end{equation}
where $V_{\rm tr}\!=\!(x^2+y^2)/2\!=\!\rho^2/2$ is the trapping potential
and $g\!=\!4\pi \gamma$ is a dimensionless coupling constant that is the
only free parameter. The conserved energy functional associated with
Eq.~(\ref{eq:1}) is given by
\begin{equation}  \label{eq:2}
W = \frac{1}{2}\, \int{\left( |\bm{\nabla} \Psi|^2 + 2 V_{\rm tr} |\Psi|^2
+ g\, |\Psi|^4 \right) dV}
\end{equation}
and yields energy in units
of $\gamma_\perp (\hbar\omega_\perp)$ where
$\gamma_\perp\!=\! \nu a_\perp \!=\! \gamma (a_\perp/a)$.

We have also considered various forms of dissipative dynamics, mainly as
relaxation algorithms for the calculation of stationary states. The simplest
possibility is to make the replacement
$i\,\partial\Psi/\partial t \to (i\!-\!\zeta) \partial\Psi/\partial t$
in Eq.~(\ref{eq:1}), where $\zeta$ is some dissipation constant.
An extreme special case widely employed in numerical simulations \cite{feder}
is the fully-dissipative dynamics  obtained by the replacement
$i\,\partial\Psi/\partial t \to -\partial\Psi/\partial t$.
However, in either case, particle number conservation is violated
and special precautions are necessary to enforce a definite number
of particles.

An interesting alternative that preserves particle number was originally
introduced by Carlson \cite{carlson} for the study of the homogeneous BEC
and is simply adapted to the case of a confined BEC. Actually, we have
employed a special case of Carlson's dissipative dynamics governed by
\begin{equation}  \label{eq:3}
i\, \frac{\partial\Psi}{\partial t} 
- \zeta\, \Psi \frac{\partial}{\partial t} |\Psi|^2 =
-\frac{1}{2} \Delta\Psi + V_{\rm tr} \Psi + g |\Psi|^2 \Psi\,.
\end{equation}
An immediate consequence of Eq.~(\ref{eq:3}) is the continuity equation
\begin{equation}  \label{eq:4}
\frac{\partial n}{\partial t} + \bm{\nabla} \cdot \bm{J} = 0\,,
\end{equation}
irrespective of the specific value of the dissipation constant $\zeta$.
Here, $n\!=\!|\Psi|^2$ is the local particle density and
\begin{equation}  \label{eq:5}
 \bm{J} = \frac{1}{2 i}\, (\Psi^* \bm{\nabla}\Psi - \bm{\nabla}\Psi^* \Psi)
        = n \bm{\nabla} \phi
\end{equation}
is the usual current density, where $\phi$ is the phase of the wave function
$\Psi \!=\! \sqrt{n}\, e^{i\phi}$. Therefore, the total number of particles
is indeed conserved by the evolution equation (\ref{eq:3}).
However, the total energy defined from Eq.~(\ref{eq:2}) dissipates according to
\begin{equation}  \label{eq:6}
\frac{d W}{d t} = - \zeta\, \int{\left(\frac{\partial n}{\partial t}\right)^2
dV},
\end{equation}
as long as the particle density $n$ remains time dependent.

Although both types of dissipation discussed above are 
employed in our numerical calculations, Carlson's proposal
has some advantages for our current purposes.
We have thus developed a discrete version of Eq. (3) on a cubic
lattice to test the stability of solitons and eventually obtain
rough estimates of other possible stationary states. These
estimates are then fed into a Newton-Raphson algorithm of the type
considered in Ref.~\cite{komineas2} to obtain our final results more
accurately as well as efficiently. The latter algorithm is currently upgraded
to a fully-3D iterative scheme that does not necessarily enforce
axial symmetry. Explicit results are discussed in the following
sections.

\section{static solitary waves}

We first consider a special class of stationary states obtained by making
the substitution $\Psi(\bm{r},t) \to e^{-i\mu t} \Psi(\bm{r})$ in
Eq.~(\ref{eq:1}) to derive the static equation
\begin{equation}  \label{eq:7}
\mu \Psi =  -\frac{1}{2} \Delta\Psi + \frac{1}{2} \rho^2\Psi
+ g |\Psi|^2 \Psi\,,
\end{equation}
where $\mu$ is the chemical potential (in units of $\hbar\omega_\perp$)
required to enforce a definite number of atoms. The simplest and
most important solution of Eq.~(\ref{eq:7}) is the ground state wave
function $\Psi\!=\!\Psi_0(\rho)$ which depends only on the radial distance
$\rho\!=\!(x^2+y^2)^{1/2}$ and may be chosen to be real and positive.
It also satisfies the constraint
\begin{equation}  \label{eq:8}
\int_0^\infty{2\pi \rho d\rho\, \Psi_0^2(\rho)} = 1\,,
\end{equation}
by a specific choice of the chemical potential $\mu\!=\!\mu(\gamma)$ for
each value of the dimensionless coupling constant $\gamma$.
Explicit illustrations of the ground state may be found in
Ref.~\cite{komineas2}.

A more general class of excited self-consistent states is obtained by
solving Eq.~(\ref{eq:7}) with the same chemical potential, as is appropriate
on an infinitely elongated trap, and boundary conditions
\begin{equation}  \label{eq:9}
\lim_{z\to \pm \infty}  |\Psi(x,y,z)| = |\Psi_0(\rho)|, \quad
\lim_{z\to \pm \infty} \frac{\partial\Psi}{\partial z} = 0\,.
\end{equation}
The class of solutions actually discussed in the present paper is
further restricted by the parity relation
\begin{equation}  \label{eq:10}
\Psi(x,y,-z) = \Psi^*(x,y,z)\,,
\end{equation}
modulo an overall constant phase  that is set equal to zero.
Relation (\ref{eq:10}) is obviously consistent with Eq.~(\ref{eq:7}).

To conclude the general description, we also consider the excitation
energy defined as
\begin{equation}  \label{eq:11}
E = W - W_0 - \mu\,\int(|\Psi|^2-|\Psi_0|^2)\, dV\,,
\end{equation}
where both $W$ and $W_0$ are calculated from the energy functional (\ref{eq:2})
applied for a general self-consistent state $\Psi$ and the ground state
$\Psi_0$, respectively. The presence in Eq.~(\ref{eq:11}) of a term
that is proportional to the chemical potential $\mu$ ensures that we
are calculating the energy difference between two states with the same
number of particles.

\begin{figure}
  \epsfig{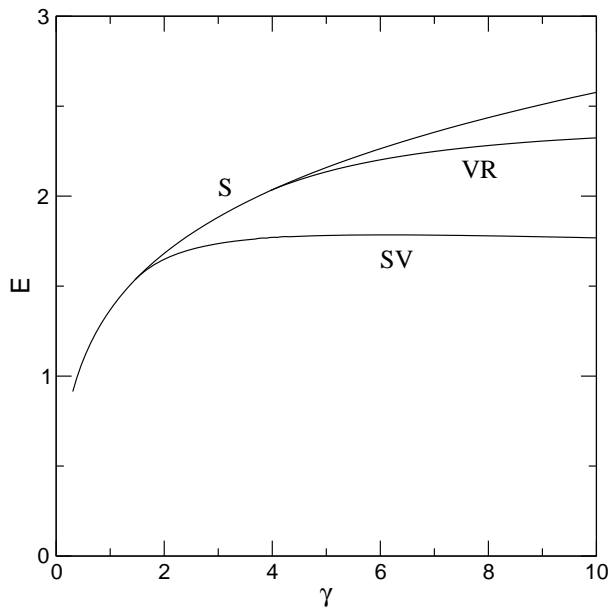}
  \caption{Excitation energy $E$ in units of $\gamma_\perp (\hbar\omega_\perp)$
as a function of the dimensionless coupling constant $\gamma$,
for static solitary waves such as a soliton (S), a solitonic vortex (SV),
and a vortex ring (VR). Bifurcations occur at the two critical couplings
$\gamma_0\!=\!1.5$ and $\gamma_1\!=\!4$.}
\label{fig:1}
\end{figure}

In recent work \cite{komineas1,komineas2} we restricted ourselves to
axisymmetric states described by a wave function of the form
$\Psi\!=\!\Psi(\rho,z)$. The simplest possibility is a static (black) soliton
whose wave function is odd under the parity reflection $z\to -z$,
and purely imaginary in view of the phase convention adopted in
Eq.~(\ref{eq:10}). The excitation energy of a static soliton calculated
from Eq.~(\ref{eq:11}) is illustrated in Fig.~\ref{fig:1}
as a function of the dimensionless coupling constant $\gamma$.

When $\gamma$ exceeds the critical value $\gamma_1\!=\!4$, a new class
of axisymmetric solitary waves emerges with the standard characteristics
of vortex rings \cite{komineas1,komineas2}. Unlike the case of a
homogeneous BEC, a vortex ring may now be static and is described by
a {\it complex} wave function that satisfies the parity relation
$\Psi(\rho,-z)\!=\!\Psi^*(\rho,z)$. The calculated excitation energy of
a static vortex ring is also shown in Fig.~\ref{fig:1} and is seen to smoothly
bifurcate from the soliton branch at the critical coupling $\gamma_1$,
while it remains consistently lower than the soliton energy for
$\gamma \!>\! \gamma_1$. Therefore, one may conclude that a radial ($m\!=\!0$)
perturbation of a soliton will render it unstable when $\gamma \!>\! \gamma_1$
and eventually transform it into an axisymmetric vortex ring.

One is tempted to presume that the soliton is stable for
$\gamma \!<\! \gamma_1$.
However, the linear stability analysis of Ref.~\cite{muryshev1,feder}
indicates that the primary soliton instability sets in at a lower
critical coupling $\gamma_0 \!<\! \gamma_1$ through azimuthal ($m\!=\!1$)
perturbations. To probe for such a possibility, we solve the dissipative
Eq.~(\ref{eq:3}) with initial condition supplied by the calculated
static soliton which is found to remain stable at all times when
$\gamma\!<\! \gamma_0 \!=\! 1.5$. However,
for $\gamma \!>\! \gamma_0$, the inherent
numerical noise introduces an instability that transforms the soliton into
a non-axisymmetric structure which is subsequently rectified by
feeding it into a 3D Newton-Raphson iterative algorithm that is  designed
to directly solve the static equation (\ref{eq:7}).

\begin{figure}
  \psfig{file=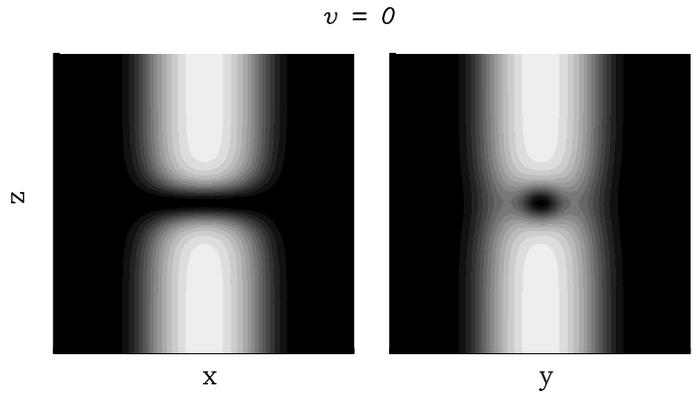,width=9cm,bbllx=65bp,bblly=425bp,bburx=520bp,bbury=680bp}
  \caption{Profile of a static ($v\!=\!0$) solitonic vortex for
$\gamma\!=\!3$, illustrated through density plots over two planes that
are perpendicular to each other and both contain the symmetry ($z$) axis
of the cylindrical trap. Regions with high particle density are
bright while regions with zero density are black. The size of each frame
is [-5,5]$\times$[-5,5].
}
\label{fig:2}
\end{figure}

We thus obtain a static solitary wave that is partially illustrated
in Fig.~\ref{fig:2} for $\gamma\!=\!3$ through density plots over two planes
that are perpendicular to each other and both contain the $z$ axis.
We have furthermore examined the phase to establish that the calculated
wave function describes a deformed vortex that pierces through the
trap in a direction that is perpendicular to its symmetry ($z$) axis.
In Fig.~\ref{eq:2}, the vortex axis is identified with the $x$ axis
as a matter of convention, whereas a degenerate class of equivalent
configurations may be obtained by simple azimuthal rotations around
the $z$ axis. The resulting solitary wave displays the characteristics of
a solitonic vortex calculated earlier by Brand and Reinhardt
\cite{brand1,brand2} in traps of varying geometry.

The energy of the solitonic vortex is depicted in Fig.~\ref{fig:1} as a
function of $\gamma$ and is seen to emerge from the soliton branch
at the critical coupling $\gamma_0\!=\!1.5$, while it remains smaller
than the energy of both the soliton and the vortex ring for all
$\gamma \!>\! \gamma_0$. The physical picture can be explained by returning
to the initial-value problem associated with the dissipative Eq.~(\ref{eq:3})
using the static soliton as initial condition. As mentioned already,
the soliton remains stable for $\gamma \!<\! \gamma_0$ but decays into a
solitonic vortex when $\gamma_0 \!<\! \gamma \!<\! \gamma_1$. The pattern is more
complex for $\gamma \!>\! \gamma_1$ where the soliton initially deforms radially
to become an axisymmetric vortex ring which eventually decays into
a stable solitonic vortex.

Therefore, a vortex ring is strictly speaking unstable. But our real-time
simulations indicate that vortex rings are sufficiently long lived to
be observed, as was actually the case in the experiment of Ref.~\cite{anderson}
-- albeit in the special limit of a spherical trap.
On the other hand, it is curious that the predicted solitonic vortex
has not yet been seen in experiment.

The picture is completed with a few remarks. One should expect that there
exists a series of additional critical couplings (densities) at which the
soliton bifurcates into multiple vortex rings and/or solitonic vortices.
For example, a double vortex ring was calculated in Ref.~\cite{komineas2}
for $\gamma \!>\! \gamma_2 \!=\! 12$. In other words, we have only described the
two basic patterns of instability that correspond to the appearance of
unstable modes in the $m\!=\!0$ and $m\!=\!1$ sectors
of the linear BdG equations. These two modes may be dominant at intermediate
densities, but more complicated patterns should be expected to occur
at higher densities, as is evident in the experiment of Ref.~\cite{anderson}
which is further discussed in Appendix B.

\section{moving solitary waves}

All three types of static solitary waves discussed in the preceding section
are special members of a more general class of stationary states that
propagate along the cylindrical trap with constant velocity $v$.
These are obtained by making the substitution
$\Psi(\bm{r},t) \to e^{-i\mu t} \Psi(x,y,\xi)$ in Eq.~(\ref{eq:1}),
with $\xi\!=\!z-v t$, to derive the stationary differential equation
\begin{eqnarray}  \label{eq:12}
\mu \Psi - i v \frac{\partial\Psi}{\partial \xi} & = &
 -\frac{1}{2} \Delta\Psi + \frac{1}{2} \rho^2\Psi + g |\Psi|^2 \Psi\,, \\
\noalign{\medskip}
 \Delta & = & \frac{\partial^2}{\partial x^2} + \frac{\partial^2}{\partial y^2}
            + \frac{\partial^2}{\partial\xi^2}\,,
\nonumber
\end{eqnarray}
which reduces to the static equation (\ref{eq:7}) at $v\!=\!0$. The boundary
conditions (\ref{eq:9}) and the parity relation (\ref{eq:10}) are also
generalized by the simple replacement $z \to \xi$.

\begin{figure}
  \psfig{file=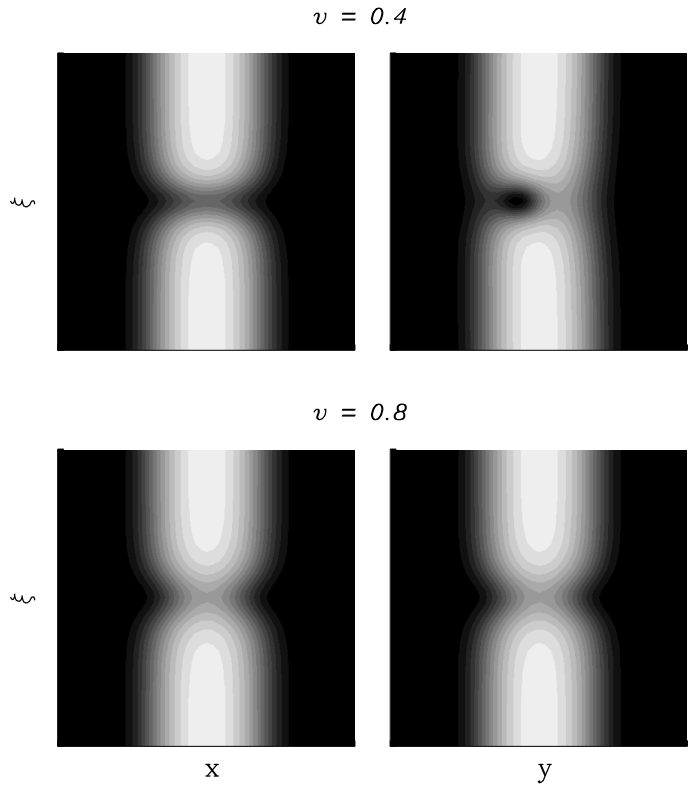,width=9cm,bbllx=60bp,bblly=150bp,bburx=520bp,bbury=670bp}
  \caption{Profile of a moving solitonic vortex for $\gamma\!=\!3$ and two
values of the velocity $v\!=\!0.4$ and 0.8. Conventions are the same
as in Fig.~\ref{fig:2} except that $\xi\!=\!z-v t$ now measures the distance
from the center of the moving solitonic vortex. The speed of sound
calculated at $\gamma\!=\!3$ is $c\!=\!1.3$.
}
\label{fig:3}
\end{figure}

Equation (\ref{eq:12}) is solved by feeding a static ($v\!=\!0$) solitary wave
into the Newton-Raphson algorithm and by upgrading the velocity typically
in steps of $\delta v \!=\! \pm 0.1$. The results for axisymmetric solitons
and vortex rings already presented in Ref.~\cite{komineas1,komineas2}
have been confirmed using the currently employed fully-3D algorithm.
Therefore, in the following, we shall mostly discuss the case of a moving
solitonic vortex for which no calculation has been given in the literature.
The static solitonic vortex of Fig.~\ref{fig:2} is significantly reshaped
when it moves, as is illustrated in Fig.~\ref{fig:3} for two values of
the velocity ($v\!=\!0.4$ and 0.8) that are both smaller than the speed of
sound $c\!=\!1.3$ calculated for $\gamma\!=\!3$ \cite{komineas2}.
The $v\!=\!0.4$ entry of Fig.~\ref{fig:3} makes it clear that a moving
solitonic vortex is shifted off center. While the vortex axis remains
within the basal $xy$ plane, it is clearly displaced away from the center
of the trap. The displacement increases with increasing speed until we
reach a critical speed $v_0\!=\!0.8$ where the vortex disappears from the
trap, in the sense that the solitary wave is converted into a gray
axisymmetric soliton. This implies, in particular, that the density plots
over the $xz$ and $yz$ planes become identical, as is apparent in the
$v\!=\!0.8$ entry of Fig.~\ref{fig:3}.

For $v_0 \!<\! |v| \!<\! c$ the calculated solitary wave is indeed identical
to a moving axisymmetric soliton and reduces to a weakly nonlinear
soundlike pulse in the limit $|v| \to c$. A corollary of this discussion
is that a high-speed soliton can be stable even for couplings $\gamma$ that
are greater than the critical coupling $\gamma_0\!=\!1.5$, in apparent
agreement with a similar conclusion reached in Ref.~\cite{muryshev2}
that is now made more precise. Specifically, for $\gamma \!<\! \gamma_0$
a solitonic vortex does not exist and the soliton is stable over the
entire range of velocities $0 \!<\! |v| \!<\! c$, as expected.
For $\gamma \!>\! \gamma_0$, the soliton is stable over the limited range
$v_0 \!<\! |v| \!<\! c$, where $v_0\!=\!v_0(\gamma)$ is the
critical speed at which the
solitonic vortex is converted into a soliton and $c\!=\!c(\gamma)$ is the
calculated speed of sound \cite{komineas2}. Clearly, $v_0(\gamma\!=\!\gamma_0)\!=\!0$,
while a numerical investigation of the ratio $v_0(\gamma)/c(\gamma)$
shows that it approaches unity with increasing $\gamma$.
However, we are not in a position to ascertain whether or not there exists
a finite (critical) $\gamma$ above which $v_0\!=\!c$ and, hence, the soliton is
not stable for any velocity. In any case, the solitonic vortex is the
fundamental mode as long as $\gamma \!>\! \gamma_0$.

Further insight into the nature of a moving solitonic vortex is obtained
by considering the behavior of its phase $\phi\!=\!\phi(x,y,\xi)$ in the
asymptotic limits $\xi \to \pm \infty$. Let
$\phi_\pm \!=\! \phi(x,y,\xi\!=\!\pm \infty)$
be the two asymptotic values of the
phase, which are, in principle, some functions of $x$ and $y$.
However, detailed numerical evidence suggests that $\phi_+$ and $\phi_-$ are,
in fact, constants that depend only on the velocity $v$.
This conclusion may be substantiated in part by an analytical argument
briefly described in Appendix A. Therefore, the asymptotic phase difference
\begin{equation}  \label{eq:13}
\delta\phi = \phi_+ - \phi_-
\end{equation}
depends only on the velocity $v$ and characterizes a solitary wave.
Incidentally, the above conclusion applies to all three types of
solitary waves discussed in the present paper and is pertinent to
their experimental production through phase imprinting
\cite{burger,denschlag}.

\begin{figure}
  \epsfig{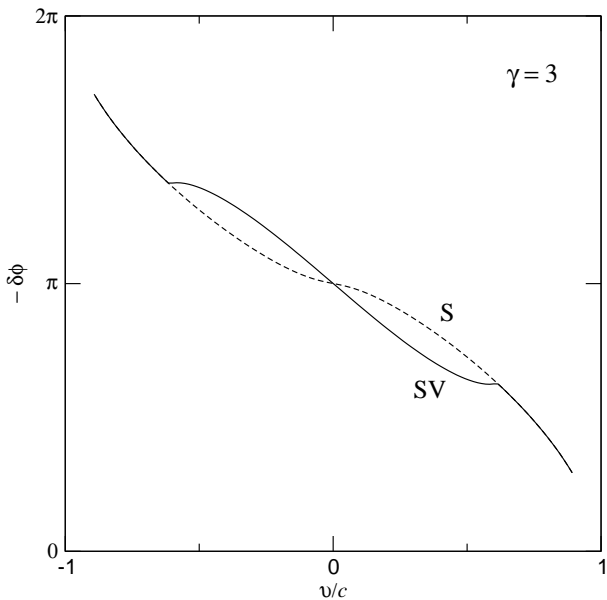}
  \caption{Asymptotic phase difference as a function of velocity
for a solitonic vortex (SV, solid line) and a soliton (S, dashed line).
The two curves merge at the critical velocities $v\!=\!\pm v_0$ where
$v_0\!=\!0.6\, c$.
}
\label{fig:4}
\end{figure}

As an illustration, we return to the solitonic vortex calculated for $\gamma\!=\!3$
and plot $-\delta\phi$ as a function of velocity by a solid line in
Fig.~\ref{fig:4}, whereas the corresponding result for an axisymmetric
soliton at $\gamma\!=\!3$ is shown by a dashed line. In both cases,
$-\delta\phi \!=\! \pi$ for $v\!=\!0$,
and $-\delta\phi \!=\! 0$ (mod $2\pi$) in the
extreme limits $v \to \pm c$ which are not actually reached in
Fig.~\ref{fig:4} due to numerical inaccuracies. The same figure
provides clear evidence for the conversion of a solitonic vortex into
an axisymmetric soliton at the critical speed $v_0\!=\!0.8$ or $v_0/c\!=\!0.6$.

The preceding discussion of the asymptotic phase is also pertinent to the
definition of the impulse of a solitary wave \cite{komineas2} which is
now simplified to
\begin{equation}  \label{eq:15}
Q = \lm - \delta\phi, \quad
\lm =  \int{n\, \frac{\partial\phi}{\partial z}\, dV}\,,
\end{equation}
where $\lm$ is the usual linear momentum and $\delta\phi$ is given by
Eq.~(\ref{eq:13}). The impulse $Q$ is measured in units of
$\hbar \nu \!=\! \gamma_\perp (\hbar/a_\perp)$.

\begin{figure}
  \epsfig{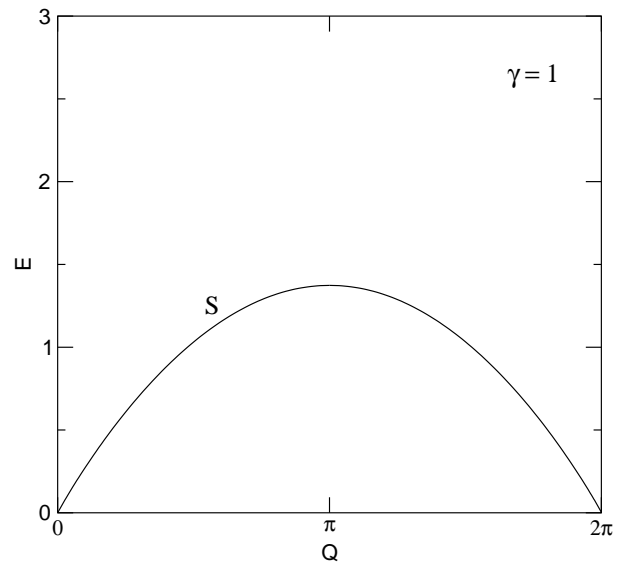}
  \caption{Excitation energy $E$ in units of $\gamma_\perp (\hbar\omega_\perp)$
vs impulse $Q$ in units of $\gamma_\perp (\hbar/a_\perp)$ for a
soliton (S) at $\gamma\!=\! 1$.
}
\label{fig:5}
\end{figure}

A solitary wave may then be characterized by the impulse $Q$ and the
excitation energy $E$ defined in Eq.~(\ref{eq:11}). Both $E$ and $Q$
are definite functions of the velocity $v$ that can be eliminated to yield
the dispersion $E\!=\!E(Q)$.
A nontrivial check of consistency is provided by the
group velocity relation $v\!=\!dE/dQ$ which was accurately confirmed by our
numerical calculation. For comparison purposes, we reproduce in
Fig.~\ref{fig:5} the dispersion of a stable axisymmetric soliton
calculated for $\gamma\!=\!1 \!<\! \gamma_0$ in Ref.~\cite{komineas2}.
This dispersion displays the characteristics of a mode originally derived
by Lieb \cite{lieb} within a full quantum treatment of a homogeneous
1D Bose gas based on the Bethe Ansatz, and was later rederived by a
semiclassical calculation based on the solitary wave of the 1D Gross-Pitaevskii
model \cite{kulish,ishikawa}. A similar mode was also derived within an
effective 1D model  that is designed to approximately describe
a cylindrical trap \cite{jackson}.

\begin{figure}
  \epsfig{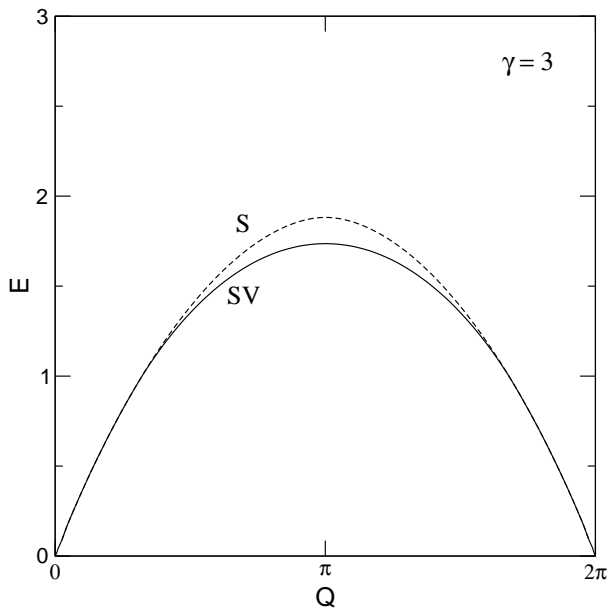}
  \caption{Excitation energy $E$ vs impulse $Q$ for a
solitonic vortex (SV, solid line) and a soliton (S, dashed line) at
$\gamma\!=\!3$. The two dispersions merge at two points that correspond
to velocities $v\!=\!\pm v_0$ where $v_0\!=\!0.6\, c$.
}
\label{fig:6}
\end{figure}

We are now in a position to calculate the dispersion of a moving solitonic
vortex. The dispersion is shown by a solid line in Fig.~\ref{fig:6} and
is seen to retain the basic characteristics of a Lieb-type mode.
We further plot by a dashed line the dispersion of an axisymmetric soliton
which is expected to be unstable at $\gamma\!=\!3$. However, the two dispersions
shown in  Fig.~\ref{fig:6} merge at two critical points that correspond
to velocities $v\!=\!\pm v_0$ where $v_0\!=\!0.6\, c$ is the critical speed at
which the solitonic vortex is converted into an axisymmetric soliton.

\begin{figure}
  \epsfig{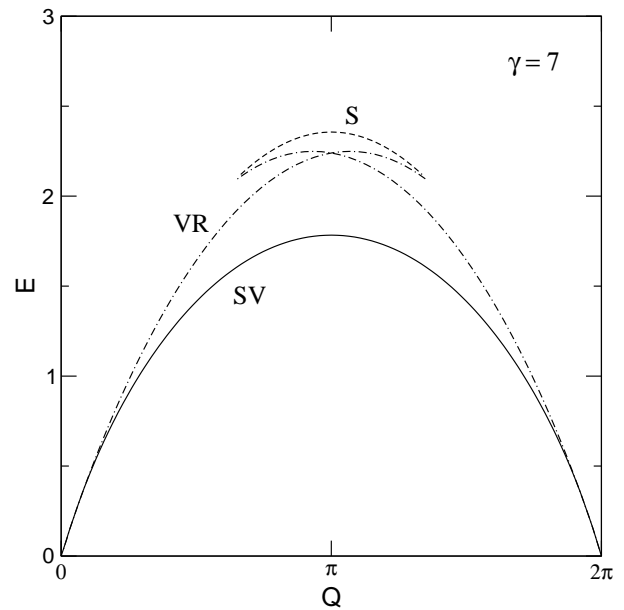}
  \caption{Excitation energy $E$ vs impulse $Q$ for a
solitonic vortex (SV, solid line), a vortex ring (VR, dotted dashed line)
and a soliton (S, dashed line) at
$\gamma\!=\!7$. The SV and VR dispersions merge at two points that
correspond to velocities $v\!=\!\pm v_0$ where $v_0\!=\!0.8 c$.
}
\label{fig:7}
\end{figure}

A subtle detail not seen in our limited illustrations is that
the calculated soliton for $\gamma\!=\!3$ exhibits some ringlike features,
when $v\neq 0$, which are not present for $v\!=\!0$ and also disappear before
reaching the critical speed $v_0$. These features could be viewed as
preliminary signals for the formation of fully-fledged vortex rings
that occur for stronger couplings in the region
$\gamma \!>\! \gamma_1 \!=\! 4$.

Since a detailed calculation of both static and moving vortex rings was already described in Ref.~\cite{komineas1,komineas2}, we merely complete the picture
by illustrating in Fig.~\ref{fig:7} the dispersions of a soliton,
a vortex ring, as well as a solitonic vortex calculated for
$\gamma\!=\!7$ where
all three possibilities occur according to the bifurcation pattern
of Fig.~\ref{fig:1}. The solitonic vortex is still characterized by
a Lieb-type dispersion that merges with the dispersion of the vortex ring
at the critical speed  $v_0\!=\!0.8\, c$,
where $c\!=\!1.6$ is the speed of sound
calculated at $\gamma\!=\!7$. For $v_0 \!<\! |v| \!<\! c$
both the solitonic vortex and
the vortex ring become identical to a gray axisymmetric soliton.
In fact, the vortex ring transforms into a soliton at a speed somewhat
smaller than $v_0$.

Therefore, it is fair to say that the primary mode associated with the
snake instability of a soliton is indeed the solitonic vortex
\cite{brand1} whose calculation has been significantly extended in the
present paper to account for steady motion along a cylindrical trap.
Actually, one may envisage a further generalization to a spinning
solitonic vortex that undergoes rigid-body precession around the $z$ axis
at constant frequency $\omega$, while it moves along the same axis
at constant velocity $v$. One could then calculate the excitation energy
$E\!=\!E(\omega,v)$, the impulse $Q\!=\!Q(\omega,v)$, and the azimuthal
angular momentum $L_z\!=\!L_z(\omega,v)$ to eventually derive a dispersion
of the form $E\!=\!E(Q,L_z)$. The Lieb-type dispersion of a solitonic vortex
shown by a solid line in Figs.~\ref{fig:6} and \ref{fig:7} would then
be the lower edge of a continuum of states parametrized by the angular
momentum $L_z$. This interesting possibility will not be discussed
further in the present paper.

\section{conclusion}

A phase imprint $\delta\phi$ applied between the two ends of an elongated
trap leads to the formation of pure quasi-1D solitons only for weak
couplings (densities) in the region $\gamma \!<\! \gamma_0 \!=\! 1.5$.
For intermediate couplings in the region $\gamma_0 \!<\! \gamma \!<\! \gamma_1 \!=\! 4$,
solitons may be produced temporarily but they eventually decay into
stable solitonic vortices. A new pattern emerges for $\gamma \!>\! \gamma_1$
where a phase imprint should initially produce unstable axisymmetric
solitons that subsequently deform radially to become axisymmetric
vortex rings and finally decay into solitonic vortices.
This pattern is further complicated by the fact that high-speed vortex
rings or solitonic vortices are indistinguishable from axisymmetric
solitons.

It is thus important to reexamine the parameters of the experiment described
in Ref.~\cite{burger}. According to the estimates obtained in Appendix B,
the length of the trap is significantly larger than its radius.
Hence the approximation by an infinite cylindrical trap may be reasonable.
Furthermore, a conservative estimate of the effective coupling leads to
$\gamma \sim 7$ which lies in the region $\gamma \!>\! \gamma_1$.
Therefore, all three types of solitary waves theoretically discussed
in the present paper should have been produced in the original experiment
of Ref.~\cite{burger}. Also note that non-stationary vortex rings
were actually detected in the experiment of Ref.~\cite{anderson} but
on a spherical trap that is significantly different than the
cylindrical trap considered here. Nevertheless, our present
analysis can be extended to the case of a finite trap, as discussed
in Appendix B.

Finally, the calculation of the present paper clearly demonstrates that
a Lieb-type mode appears in the excitation spectrum for all densities,
either in the form of a quasi-1D soliton (for $\gamma \!<\! \gamma_0$)
or in the form of a solitonic vortex (for $\gamma \!>\! \gamma_0$).
It is thus interesting to examine whether or not such a mode can be measured.

\begin{acknowledgments}
We are grateful to J. Brand for a stimulating discussion that brought
the work of Ref.~\cite{brand1} to our attention, to N.R. Cooper
for his constant interest in the present effort, and to
A.S. Fokas, and N.S. Manton for a number of useful comments.
This work was supported by the EPSRC under contract GR/R96026/01 (S.K.).
\end{acknowledgments}

\begin{appendix}
\section{ASYMPTOTIC PHASE}

The essence of the analytical argument given below was suggested to us by
Fokas \cite{fokas}. For a solitary wave traveling with constant velocity $v$,
the continuity equation (\ref{eq:4}) may be written as
$-v {\partial n}/{\partial \xi}
+ \bm{\nabla}\cdot(n \bm{\nabla}\phi) \!=\!0.$
In the limits $\xi\! \to\! \pm \infty,\,
 n\! \to\! n_0(\rho) \!=\! |\Psi_0(\rho)|^2$
and all $\xi$ (or $z$) derivatives vanish on account of the boundary
conditions (\ref{eq:9}). Hence,
$\bm{\nabla}_\perp\cdot [n_0(\rho) \bm{\nabla}_\perp \phi_\pm(x,y)] \!=\! 0$,
where $\bm{\nabla}_\perp$ is the 2D gradient operator and
$\phi_\pm(x,y) \!=\! \phi(x,y,\xi\!=\!\pm\infty)$
are the unknown asymptotic limits
of the phase. If we further employ cylindrical coordinates
($x\!=\!\rho \cos\!\varphi, y \!=\! \rho\sin\!\varphi$)
the asymptotic form of the
continuity equation reads 
\begin{equation}  \label{eq:a1}
\rho\, \frac{\partial}{\partial \rho} \left[ \rho\, n_0(\rho)\,
 \frac{\partial \phi_\pm}{\partial \rho} \right]
 +  n_0(\rho)\,\frac{\partial^2 \phi_\pm}{\partial \varphi^2} = 0\,.
\end{equation}
For axisymmetric solitons and vortex rings $\phi_\pm = \phi_\pm (\rho)$
and thus the second term in Eq.~(\ref{eq:a1}) vanishes.
A first integral is then given by $d\phi_\pm/d\rho \!=\! c_1/\rho\, n_0(\rho)$
where $c_1$ is some integration constant. Also note that the energy
functional (\ref{eq:2}) contains a term $n (\partial\phi/\partial\rho)^2$
that becomes $n_0(\rho) (d\phi_\pm/d\rho)^2 \!=\! c_1^2/\rho^2 n_0(\rho)$ at the
two ends of the trap.
Since the ground-state density $n_0(\rho)$ is finite at $\rho\!=\!0$
and vanishes in the limit $\rho \to \infty$, the above term would lead
to a badly divergent integral over the radial distance $\rho$ unless
the integration constant $c_1$ vanish. Hence, for an axisymmetric
solitary wave with finite excitation energy,
$d\phi_\pm/d\rho\!=\!0$ and thus the asymptotic phases
$\phi_\pm$ can depend only on the velocity $v$.

On the other hand, we have not been able to extend the above argument
to the case of non-axisymmetric solitary waves for which both terms in
Eq.~(\ref{eq:a1}) are potentially relevant. Nevertheless, our numerical
results for the solitonic vortex discussed in the main text again suggest
that the asymptotic phases $\phi_\pm\!=\!\phi_\pm(x,y)$ are, in fact,
independent of $x$ and $y$.

\section{FINITE TRAPS}

A finite harmonic trap with axial symmetry is characterized by the two
confinement frequencies $\omega_\perp$ and $\omega_\|$.
We now introduce rationalized quantities that are in some respects
different than those employed for the discussion of the infinite cylindrical
trap. Time and distance are still measured in units of $1/\omega_\perp$
and $a_\perp$ but the wave function is rescaled according to
$\Psi \to \sqrt{N} \Psi /a_\perp^{3/2}$, where $N$ is the total number of
atoms, and thus satisfies the constraint $\int{|\Psi|^2 dV} \!=\! 1$.
The Gross-Pitaevskii equation is then formally identical to
Eq.~(\ref{eq:1}) but
\begin{equation}  \label{eq:b1}
 g = 4 \pi \alpha, \quad V_{\rm tr} = \frac{1}{2}\,(\rho^2 + \beta^2 z^2)\,,
\end{equation}
where $\alpha \!=\! N a/a_\perp$ and $\beta\!=\! \omega_\|/\omega_\perp$ are now
the two independent dimensionless parameters. To make contact with the
infinite trap, we calculate the average linear density from
$\nu(z) \!=\! (N/a_\perp)\, \int{|\Psi|^2 dx dy}$
and define the dimensionless quantity $\gamma(z) = a \nu(z)$ or
\begin{equation}  \label{eq:b2}
  \gamma(z) = \alpha \int{|\Psi|^2\, dx dy}\,,
\end{equation}
which is the analog of the dimensionless coupling constant $\gamma$ employed
in the main text, except that it now depends on $z$. Nevertheless,
$\gamma(z)$ becomes increasingly uniform in the limit $\alpha \to \infty$
and $\beta \to 0$, holding $\gamma(0) \equiv \gamma$
fixed, which is the ideal limit
of an infinite cylindrical trap studied in the main body of the paper.

As a simple illustration, we consider the Thomas-Fermi (TF) approximation
of the ground state \cite{baym} to find that
\begin{equation}  \label{eq:b3}
\gamma(z) = \frac{1}{16}\, (2 \mu - \beta^2 z^2)^2\,,
\end{equation}
where the chemical potential is given by $2 \mu \!=\! (15\alpha\beta)^{2/5}$.
Therefore, the estimated maximum value of $\gamma$ is given by
$\gamma_{\rm max} \!=\! \gamma(0) \!=\! \mu^2/4$. A more conservative estimate
is the average value defined from $\gamma_{\rm av}\!=\!\alpha/L$ where
$L\!=\!2 \sqrt{2\mu}/\beta$ is the TF length of the trap and thus
$\gamma_{\rm av}\!=\!(8/15) \gamma_{\rm max}$.
The TF radius of the trap is accordingly given by $R\!=\!\sqrt{2\mu}$.
For the parameters of the experiment described in Ref.~\cite{burger}
we find $R\!=\!2\,\mu\, \hbox{m}$ and $L\!=\!115\, \mu\, \hbox{m}$,
while $\gamma_{\rm max}\!=\!13$
and $\gamma_{\rm av}\!=\!7$. These estimates are further discussed
in the concluding section in the main text.

Now, {\it static} solitons, solitonic vortices, and vortex rings exist
also in a finite trap and we have calculated them explicitly for
various values of $\alpha$ and $\beta$ to confirm the general picture
described in Sec.~III. Here, we briefly summarize the basic facts.
For each value of the aspect ratio $\beta$, there exist two critical couplings
$\alpha_0\!=\!\alpha_0(\beta)$ and $\alpha_1 \!=\! \alpha_1(\beta)$ that
lead to a bifurcation pattern similar to that shown in
Fig.~\ref{fig:1}. The critical coupling $\alpha_0$ coincides with the
coupling where a purely imaginary eigenvalue first appears in the
$m\!=\!1$ sector of the linear BdG equations \cite{muryshev1}.
Similarly, the critical coupling $\alpha_1$ is related to an instability
that occurs in the $m\!=\!0$ sector, but the picture is now significantly
complicated by the appearance of {\it complex} modes \cite{feder}.
Our numerical solution of the BdG equations shows that there exists
an intermediate critical coupling $\alpha_1' \!<\! \alpha_1$ where two real
eigenvalues with $m\!=\!0$ coalesce and eventually become complex for
$\alpha \!>\! \alpha_1'$. The precise behavior of the complex eigenvalues
with increasing $\alpha$ depends on the specific value of $\beta$.
A common feature is that complex eigenvalues as such again disappear
before reaching the critical coupling $\alpha_1$.
The latter is characterized by the fact that only one purely imaginary
eigenvalue persists for $\alpha \!>\! \alpha_1$ and signals a new bifurcation
that leads to the appearance of axisymmetric vortex rings.

\begin{figure}
  \psfig{file=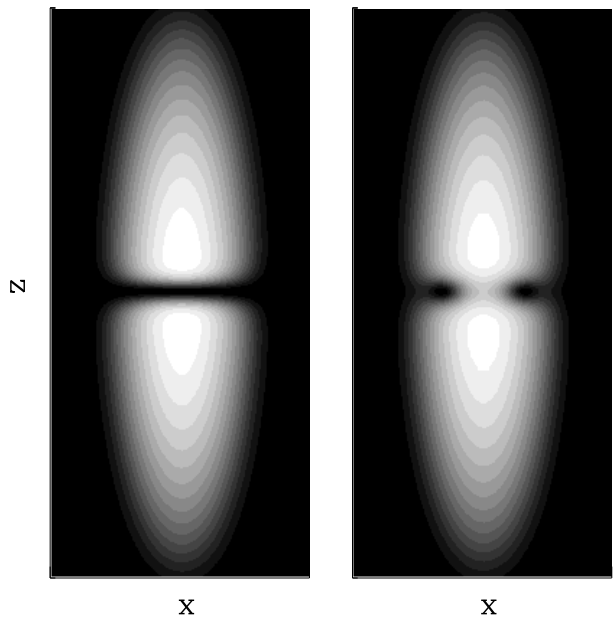,width=8cm,bbllx=110bp,bblly=360bp,bburx=465bp,bbury=720bp}
  \caption{Profiles of a static soliton (left panel) and a static
vortex ring (right panel) on a finite trap with $\alpha\!=\! 100$ and
$\beta\!=\! 1/4$, illustrated through density plots over a plane that
contains the symmetry ($z$) axis and cuts across the axisymmetric trap.
The complete pictures can be obtained by simple revolution around the
$z$ axis because both the soliton and the vortex ring are axisymmetric.
}
\label{fig:8}
\end{figure}

\begin{figure}[!t]
  \psfig{file=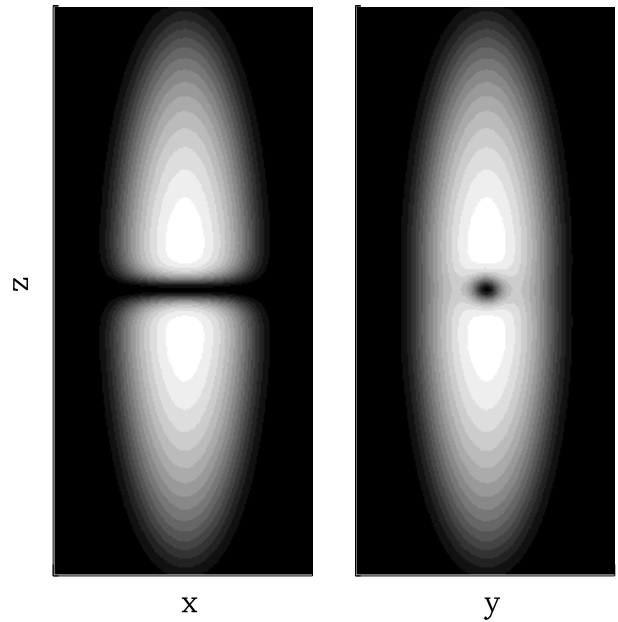,width=8cm,bbllx=110bp,bblly=360bp,bburx=465bp,bbury=725bp}
  \caption{Profile of a static solitonic vortex on a finite trap
with $\alpha\!=\! 100$ and $\beta\!=\! 1/4$, illustrated through density
plots over two planes that are perpendicular to each other and both
contain the symmetry ($z$) axis. Note that the solitonic vortex is not
axisymmetric.
}
\label{fig:9}
\end{figure}

As an example we consider the popular ratio $\beta\!=\!1/4$ for which we find
$\alpha_0 \!=\! 10, \alpha_1' \!=\! 32, \alpha_1 \!=\! 39$.
A static soliton exists for all $\alpha$ but is stable only for
$\alpha\!<\! 10$, while a static solitonic vortex appears as the most
stable structure for $\alpha\!>\! 10$. A new threshold occurs at
$\alpha\!=\! 39$ where a static vortex ring emerges with energy intermediate
between that of a soliton and a solitonic vortex, in analogy with the
situation on an infinitely cylindrical trap demonstrated in Fig.~\ref{fig:1}.
Explicit illustrations of the currently calculated profiles are given
in Figs.~\ref{fig:8},\ref{fig:9} for $\alpha\!=\!100$ and
$\beta\!=\!1/4$ where all three types of static solitary waves are possible.
For smaller $\beta$ the
critical couplings $\alpha_0$ and $\alpha_1$ increase indefinitely.
But the corresponding effective couplings $\gamma_0$ and $\gamma_1$,
numerically calculated from Eq.~(\ref{eq:b2}) applied for $z\!=\!0$,
converge to the asymptotic values $\gamma_0\!=\!1.5$ and $\gamma_1\!=\!4$ that
are in agreement with the critical couplings obtained in the main text
for an infinite cylindrical trap.
It is also interesting to consider the special limit
of a spherical ($\beta\!=\!1$)
trap in view of the experiment described in Ref.~\cite{anderson}.
The critical coupling $\alpha_0$ vanishes at $\beta\!=\!1$ and thus the
black soliton is never stable on a spherical trap. On the other hand,
the solitonic vortex now becomes degenerate with the ordinary vortex
and is stable for all $\alpha \!>\!0$. The vortex-ring threshold is here
predicted to occur at $\alpha_1\!=\!5$. For the parameters of the actual
experiment \cite{anderson} we find $\alpha\!=\!430$ which is substantially
larger than $\alpha_1$ and eventually explains the observation of
multiple vortex rings.

\end{appendix}


\end{document}